\documentstyle[twoside,fleqn,npb,epsfig]{article}         
\def\lapproxeq{\lower .7ex\hbox{$\;\stackrel{\textstyle                                              
<}{\sim}\;$}}                                              
\def\gapproxeq{\lower .7ex\hbox{$\;\stackrel{\textstyle                                              
>}{\sim}\;$}}        
\newcommand{\AmS}{{\protect\the\textfont2         
  A\kern-.1667em\lower.5ex\hbox{M}\kern-.125emS}}         
         
\title{Update of MRST partons}         
         
\author{A.D.~Martin\address{University of Durham, DH1 3LE, United Kingdom.},         
R.G.~Roberts\address{Rutherford Appleton Laboratory, Oxon, OX11 0QX, United         
Kingdom.}, W.J.~Stirling$^{\rm a}$ and R.S.~Thorne\address{Jesus College,         
University of Oxford, OX1 3DW, United Kingdom.}}         
         
\begin{document}         
         
\begin{abstract}         
We discuss topical issues concerning the MRST parton analysis.  We introduce some         
new parton sets in order to estimate the uncertainty in the cross sections predicted for         
$W$ and $Z$ production at the Tevatron and the LHC.  We compare with CTEQ5         
partons.        
\end{abstract}         
         
\maketitle         
         
MRST is a standard next-to-leading order global analysis \cite{MRST} of deep         
inelastic and related hard scattering data.  The partons were presented at DIS98 and         
the scheme dependence and higher twist effects studied in a subsequent analysis         
\cite{MRST1}.  Here we concentrate on topical issues.        
        
\section{Light quark distributions $u, d, s$}         
         
$u$ and $d$ are pinned down by DIS data, with the slope of $d/u$ also being         
constrained for $x \lapproxeq 0.3$ by the CDF $W^\pm$ rapidity asymmetry data.  In         
fact $d/u$ is determined up to $x \sim 0.7$ by the NMC $F_2^n/F_2^p$ data, {\it         
provided} that we assume that there are no deuterium binding corrections for $x >         
0.3$.  Yang and Bodek \cite{YB} included such a binding correction and described         
the data with modified partons which satisfy $d/u \rightarrow 0.2$ as $x \rightarrow         
1$, whereas MRS(T) and CTEQ parametrizations have $d/u \rightarrow 0$.  We         
repeated the global analysis, forcing $d/u \rightarrow 0.2$ and found an equally good         
fit, apart from the description of NMC $F_2^n/F_2^p$ for $x \gapproxeq 0.5$.  To fit         
these latter (uncorrected) data we require $d/u \rightarrow 0$, see also ref.~\cite{B}.          
We note, that as statistics improve, the HERA charge-current DIS data will directly         
constrain $d/u$ at large $x$.  So far the preliminary data and the MRST prediction are         
in agreement.        
        
$\bar{d}/\bar{u}$ is determined for $0.05 \lapproxeq x \lapproxeq 0.3$ by the E866         
$pp, pn \rightarrow \mu^+ \mu^- X$ data, and also by the HERMES data for $ep, en         
\rightarrow e \pi^\pm X$.  The MRST partons are in agreement with both the new         
E866 \cite{E866} and HERMES \cite{HERMES} data.        
        
In principle $\nu$ and $\bar{\nu}$ DIS data can determine $s$ and $\bar{s}$ separately.      
However we use the CCFR dimuon data to determine $s + \bar{s}$.  To quantify the     
uncertainty we obtained two new sets of MRST partons, $s\uparrow$ and $s\downarrow$     
corresponding to $(s + \bar{s})/(\bar{d} + \bar{u}) = 0.5 \pm 0.06$ at $Q^2 = 1$~GeV$^2$,     
which embrace the spread of these data.        
         
\section{Heavy quark distributions $c, b$}         
        
For a heavy quark, $H = c$ (or $b$) we should match, at $Q^2 = m_H^2$, the fixed         
flavour number scheme, describing $\gamma g \rightarrow H\bar{H}$ with $n_f =         
3(4)$, to the variable flavour number scheme with $n_f + 1$ flavours for $Q^2 >         
m_H^2$.  MRST use the Thorne-Roberts prescription \cite{TR} in which $F_2^H$         
and $\partial F_2^H/\partial \ln Q^2$ are required to be continuous at $Q^2 =         
m_H^2$.  (CTEQ use an alternative, ACOT, prescription.)~~In this way the charm         
distribution is generated perturbatively with the mass of the charm quark $m_c$ as the         
only free parameter.  Not surprisingly, due to the $g \rightarrow c\bar{c}$ transition,         
the charm distribution mirrors the form of the gluon distribution.  Again, we have         
obtained two new MRST sets, $c\uparrow$ and $c\downarrow$, corresponding to         
$m_c = 1.35 \pm 0.15$~GeV.        
         
\section{The gluon distribution $g$}        
        
In principle many processes are sensitive to the gluon distribution, {\it but} it is still         
difficult to nail down for $x \gapproxeq 0.2$ where it becomes increasingly small.  In         
this region MRST used the WA70 prompt photon data to determine the gluon.  However         
this process suffers from scale dependence (reflecting the higher order         
corrections), effects of intrinsic $k_T$ (needed to describe the E706 prompt photon data) and 
uncertainties due to fragmentation effects and isolation criteria.  To explore an acceptable 
spread of gluon distributions at large $x$, MRST presented three sets ($g\uparrow$, default 
MRST set, $g\downarrow$) corresponding to average intrinsic $k_T = 0,~0.4$ and         
0.64~GeV respectively for the WA70 data.  The arrows indicate the relative sizes of         
the gluons in the large $x$ region ($x \gapproxeq 0.2$).  The corresponding $\langle         
k_T \rangle$ needed to obtain a good description of the E706 prompt photon data is         
about 1~GeV.        
        
Due to the uncertainties associated with prompt photon production, the recent CTEQ5         
analysis \cite{CTEQ5} omits these data and instead determines the large $x$         
behaviour of the gluon using the single jet inclusive $E_T$ distributions measured at         
the Tevatron.  The above three MRST sets of partons describe the shape of the jet $E_T$         
distributions, but need to be renormalized upwards by 7\%, 13\%, 17\% respectively to         
give the correct normalization, with $g\uparrow$ being closest to CTEQ5.        
        
Turning now to the small $x$ domain, the gluon is well constrained by the observed         
behaviour of $\partial F_2/\partial \log Q^2$.  To be precise the H1 and ZEUS data determine         
the gluon for $x \lapproxeq 0.01$, and the NMC data for $x \lapproxeq 0.1$.  Inspection         
of $F_2$ at low $x$, low $Q^2$ shows that the data have flattened out, indicating a         
smaller value of $\partial F_2/\partial \log Q^2 \simeq \alpha_S (Q^2) P_{qg} \otimes         
g$.  Also $\alpha_S$ increases with decreasing $Q^2$, and as a consequence $g$         
becomes valence-like for $Q^2 \simeq 1$~GeV$^2$, which is about the lowest value of         
$Q^2$ for a satisfactory DGLAP description.        
         
\section{$W$ production at the Tevatron and LHC}         
         
The total cross sections for $W$ and $Z$ hadroproduction are known to NNLO \cite{NV}       
and the input electroweak parameters are known to high accuracy.  The main uncertainty       
comes from the input parton distributions and, to a lesser extent, $\alpha_S$ \cite{LHC}.  For     
example,       
the results for $W$ production at the Tevatron from the above sets of partons are shown in       
Fig.~1.  In addition we show the cross sections obtained from parton sets $\alpha_S       
\uparrow\uparrow$, $\alpha_S \downarrow\downarrow$ corresponding to global analyses       
with $\alpha_S (M_Z^2) = 0.1175 \pm 0.05$, and from parton sets $q\uparrow,       
q\downarrow$ corresponding to global fits with the HERA data renormalised up, down by       
2.5\%. The differences coming from $s\uparrow$, $s\downarrow$ and from $c\uparrow$,         
$c\downarrow$ are very small (and so the predictions are not shown in Fig.~1), since in the     
evolution up to $Q^2 = M_W^2$ the transitions $g \rightarrow s\bar{s}, c\bar{c}$ wash out     
the dependence on the $s, c$ input.        
        
\begin{figure}[tb]    
\begin{center}    
\mbox{\epsfig{figure=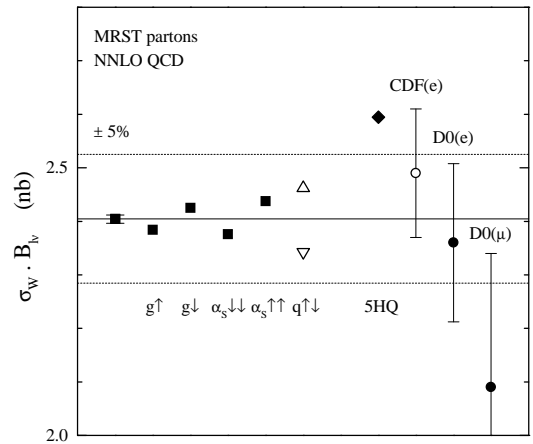,width=8cm}}    
\end{center}    
\vspace{-1truecm}            
\caption{NNLO \protect\cite{NV} predictions of the $W$ production     
cross section at the Tevatron from seven different sets of NLO MRST partons,     
and from CTEQ5HQ partons.  Also shown are the cross sections measured by the CDF and   
D0 collaborations.}       
\label{fig:largenenough}       
\end{figure}    
    
We see that the values from all the MRST sets are within about $\pm 2.5\%$ of the MRST         
default set.  The bounds are actually given by the $q\uparrow$ and $q\downarrow$ sets.  At         
first sight we might anticipate these bounds would be $\pm 5\%$ since $q\bar{q}$ initiates         
$W$ production.  However the difference due to using $q\uparrow$ and $q\downarrow$ is         
diluted by the evolution from $Q^2$ (HERA data) $\rightarrow Q^2 = M_W^2$.    
        
We also show in Fig.~1 the prediction obtained using CTEQ5HQ partons.  Surprisingly it         
differs by about 8\% from that of MRST.  Since we hope that the rate of $W$         
production is accurately predicted at the Tevatron and at the LHC, it is crucial to         
understand the origin of this difference.        
        
\section{Comparison of MRST and CTEQ5}        
        
Differences between the MRST and CTEQ partons can arise from       
\begin{itemize}       
\item the data sets included in the global fit, the $Q^2, W^2 \ldots$ cuts used,       
\item the normalization of the data sets,       
\item the treatment of heavy flavors (massless, TR and ACOT schemes),       
\item differences in the evolution codes.       
\end{itemize}       
Inspection of the resulting partons show that CTEQ5 has, first, a larger gluon at large $x$        
than MRST (due to the different data fitted), second, a larger charm distribution,         
and finally a smaller $d/u$ at very large $x$.        
        
The unexpectedly large difference between the values of the $W$ cross section calculated        
from the MRST and CTEQ5 partons, seen in Fig.~1, appears to be due to the more rapid        
evolution of the sea quarks of CTEQ5.  A similar effect was found in a comparison of various        
NLO evolution codes made during the 1996 HERA workshop \cite{CODE}.  Neither CTEQ,        
nor MRS, agreed completely with the standard code after evolution up to $Q^2 =        
100$~GeV$^2$.  The CTEQ sea was some 4\% high and the MRS sea was 1\% low and the        
gluon about 1.5\% low.  A careful check of the MRS code has recently revealed a small error        
in the NLO contribution to the evolution of the gluon.  After this correction the agreement        
between MRS and the standard evolution code is exact.  The corrected code increases the   
quarks in the relevant $x$ and $Q^2$ range by typically 1\%, and hence the MRST values of   
the $W$ cross section at the Tevatron shown in Fig.~1 are increased by about  2\%, so that  
\begin{equation}  
B_W \sigma_W \; = \; 2.45 \pm 0.06~{\rm nb}.  
\end{equation}  
A full discussion will be given in ref.~\cite{LHC}.   
       
\section{Outstanding problems}       
       
At present the two major deficiencies in the parton determinations are the absence of a        
complete NNLO analysis and the need to determine realistic errors on the partons.  This last        
task is especially problematic.  Determinations in certain $(x, Q^2)$ domains often rely       
dominantly on one experiment.  Moreover we need some way to quantify the uncertainties       
due to $k_T$ smearing, heavy target and other nuclear corrections, higher twist cuts etc.        
However, as evidenced by the contributions at this conference, continued progress is being       
made in improving our knowledge of partons.  
      
\medskip 
\noindent {\bf Acknowledgements} 
 
\medskip 
We thank Max Klein and Johannes Bl\"{u}mlein for their efficient organization of DIS99,  
 and the EU Fourth Framework Programme `Training and Mobility of 
Researchers', Network `QCD and the Deep Structure of Elementary Particles', contract 
FMRX-CT98-0194 (DG 12-MIHT) for support.

\end{document}